\documentclass[aps,prx,reprint,nofootinbib]{revtex4-1}

\usepackage{amsmath}
\usepackage{amssymb}	
\usepackage{bbold}
\usepackage{bm} 

\usepackage{xcolor}
\definecolor{myBlue}{RGB}{50,117,180}
\definecolor{myRed}{RGB}{200,40,40}
\definecolor{myGreen}{RGB}{34,139,34}
\definecolor{myLightBlue}{RGB}{193,223,255}
\definecolor{myOrange}{RGB}{255,185,0}

\usepackage{cancel} 
\usepackage[normalem]{ulem} 
\usepackage{changepage}
\usepackage{float}
\usepackage{mathrsfs}
\usepackage{url}
\usepackage{xfrac}
\usepackage{graphicx}
\usepackage{lipsum} 
\graphicspath{{/}}

\newcommand{\comment}[1]{\ignorespaces} 

\renewcommand{\v}[1]{\ensuremath{\mathbf{#1}}} 
 
\renewcommand{\d}[2]{\frac{d #1}{d #2}} 
 
\providecommand{\e}[1]{\ensuremath{\times 10^{#1}}}
\DeclareMathAlphabet{\mathsfit}{\encodingdefault}{\sfdefault}{m}{n}
\SetMathAlphabet{\mathsfit}{bold}{\encodingdefault}{\sfdefault}{bx}{n}
\newcommand{\ms}[1]{\bm{\mathsfit{#1}}} 
\newcommand{\ws}[1]{\mathsf{#1}} 

\makeatletter
\newcommand*\bigcdot{\mathpalette\bigcdot@{.5}}
\newcommand*\bigcdot@[2]{\mathbin{\vcenter{\hbox{\scalebox{#2}{$\m@th#1\bullet$}}}}}
\makeatother

\makeatletter
\newcommand{\raisemath}[1]{\mathpalette{\raisem@th{#1}}}
\newcommand{\raisem@th}[3]{\raisebox{#1}{$#2#3$}}
\makeatother

\usepackage{accents}

\usepackage{stackengine} 

\makeatletter 
\renewcommand*\env@matrix[1][*\c@MaxMatrixCols c]{%
  \hskip -\arraycolsep
  \let\@ifnextchar\new@ifnextchar
  \array{#1}}
\makeatother

\usepackage{pgfplots}
\usepgfplotslibrary{patchplots}
\usepgfplotslibrary{external}
\usepackage{tikz}
\usetikzlibrary{external}
\usetikzlibrary{calc}
\usetikzlibrary{arrows,shapes,backgrounds,arrows.meta,shapes.arrows,automata,quotes,intersections}
\usetikzlibrary{decorations.text}
\usetikzlibrary{decorations.fractals}
\usetikzlibrary{decorations.pathmorphing}
\usetikzlibrary{decorations.markings}

\usetikzlibrary{decorations.pathreplacing}
\makeatletter
    \let\pgf@decorate@@brace@brace@code@old\pgf@decorate@@brace@brace@code
    \def\pgf@decorate@@brace@brace@code{
        \ifdim\pgfdecoratedremainingdistance<4\pgfdecorationsegmentamplitude
            \pgftransformxscale{\pgfdecoratedremainingdistance/4\pgfdecorationsegmentamplitude}
            \pgfdecoratedremainingdistance=4\pgfdecorationsegmentamplitude
        \fi
        \pgf@decorate@@brace@brace@code@old
    }
\makeatother

\newcommand*\mZ{\mathbb{Z}}

\newcommand*\mR{\mathbb{R}}

\newcommand*\mT{\mathbb{T}}
\newcommand*\mP{\mathbb{P}}

\newcommand*\mD{\mathrm{D}}

\newcommand*\mH{\mathcal{H}}

\newcommand*\mJ{\mathcal{J}}

\newcommand*\mL{\mathcal{L}}
\newcommand*\mO{\mathcal{O}}

\newcommand*\mg{\mathfrak{g}}

\newcommand*\mmp{\mathfrak{p}}

\newcommand*\mt{\mathfrak{t}}
\newcommand*\mOne{\mathbb{1}}
\newcommand*\mZero{\mathbb{0}}
\newcommand*\mmT{\mT^{3,1}}

\newcommand*\vphi{\varphi}

\newcommand*\md{\mathrm{d}}

\newcommand*\n{[\v{n}]}

\DeclareMathAlphabet{\mathpzc}{OT1}{pzc}{m}{it}

\newcommand*\tleft{\triangleright}

\makeatletter
\newcommand*{\defeq}{\mathrel{\rlap{%
                     \raisebox{0.3ex}{$\m@th\cdot$}}%
                     \raisebox{-0.3ex}{$\m@th\cdot$}}%
                     =}
\newcommand*{\eqdef}{=\mathrel{\rlap{%
                     \raisebox{0.3ex}{$\m@th\cdot$}}%
                     \raisebox{-0.3ex}{$\m@th\cdot$}}%
                     }
\makeatother

\usepackage{mathtools}

\usepackage{hhline}

\usepackage{stmaryrd}

\usepackage{enumitem}

\usepackage{diagrams}
\usetikzlibrary{cd}

\definecolor{light-gray}{gray}{.85}
\newsavebox{\songboxbox}

\usepackage[most]{tcolorbox}
\newtcbox{\MyBox}[1][red]{on line, size=tight, boxsep=1pt, colframe=#1!50!black, colback=#1!10!white}

\usepackage[framemethod=TikZ]{mdframed}
\mdfsetup{%
    outerlinewidth=1pt,
    innertopmargin=6pt,
    innerbottommargin=6pt,
    skipabove=10pt,
    skipbelow=10pt,
    backgroundcolor=black!10,
    roundcorner=20pt
}

\DeclareFontFamily{U}{matha}{\hyphenchar\font45}
\DeclareFontShape{U}{matha}{m}{n}{
      <5> <6> <7> <8> <9> <10> gen * matha
      <10.95> matha10 <12> <14.4> <17.28> <20.74> <24.88> matha12
      }{}
\makeatletter
\newcommand{\blandor}[1]{\mathbin{\@blandor{#1}}}
\newcommand{\@blandor}[1]{\mathchoice
  {\@@blandor{#1}{\tf@size}}
  {\@@blandor{#1}{\tf@size}}
  {\@@blandor{#1}{\sf@size}}
  {\@@blandor{#1}{\ssf@size}}
}
\newcommand{\@@blandor}[2]{%
    \raisebox{.1ex}{\rotatebox[origin=c]{#1}{%
      \fontsize{#2}{#2}\usefont{U}{matha}{m}{n}\symbol{\string"CE}}}%
}
\makeatother

\usepackage{slashed} 
\newcommand{\cmmnt}[1]{\ignorespaces} 

\usepackage{environ}
\NewEnviron{eqn}{
\begin{align}\begin{split}
\BODY
\end{split}\end{align}
}

\newcommand{\latlink}{%
\begin{tikzpicture}[every node/.style={inner xsep=0pt,outer xsep=0pt}]%
\draw[fill = white] (0ex,.25ex) circle (.3ex);
\draw[thick] (.3ex,.25ex) -- (1.6ex,.25ex);%
\draw[fill = white] (1.9ex, .25ex) circle (.3ex);%
\end{tikzpicture}%
}
\newcommand{\latlinkna}{%
\begin{tikzpicture}[every node/.style={inner xsep=0pt,outer xsep=0pt}]
\node (A) at (0ex,1.0ex) {\tiny$\v{n}$};
\node (B) at (0.95ex,0.8ex) {\tiny$a$};
\draw[fill = white] (0ex,.25ex) circle (.3ex);
\draw[thick] (.3ex,.25ex) -- (1.6ex,.25ex);%
\draw[fill = white] (1.9ex, .25ex) circle (.3ex);%
\end{tikzpicture}%
}
\newcommand{\latlinkan}{%
\begin{tikzpicture}[every node/.style={inner xsep=0pt,outer xsep=0pt}]
\node (A) at (1.95ex,1.0ex) {\tiny$\v{n}$};
\node (B) at (0.95ex,0.8ex) {\tiny$a$};
\draw[fill = white] (0ex,.25ex) circle (.3ex);
\draw[thick] (.3ex,.25ex) -- (1.6ex,.25ex);%
\draw[fill = white] (1.9ex, .25ex) circle (.3ex);%
\end{tikzpicture}%
}

\newcommand{\latsquare}{\ensuremath{%
  \mathchoice{\raisebox{-.5ex}{\includegraphics[height=2.3ex]{latsquare}}}
    {\raisebox{-.5ex}{\includegraphics[height=2.3ex]{latsquare}}}
    {\raisebox{-.5ex}{\includegraphics[height=1.8ex]{latsquare}}}
    {\raisebox{-.5ex}{\includegraphics[height=1.3ex]{latsquare}}}
}}
\newcommand{\latsquarenab}{\ensuremath{%
  \mathchoice{\raisebox{-.4ex}{\includegraphics[height=2.3ex]{latsquarenab}}}
    {\raisebox{-.4ex}{\includegraphics[height=2.3ex]{latsquarenab}}}
    {\raisebox{-.4ex}{\includegraphics[height=1.8ex]{latsquarenab}}}
    {\raisebox{-.4ex}{\includegraphics[height=1.3ex]{latsquarenab}}}
}}

\usepackage{bbding}
\DeclareFontFamily{U}{MnSymbolC}{}
\DeclareSymbolFont{MnSyC}{U}{MnSymbolC}{m}{n}
\DeclareMathSymbol{\diamondplus}{\mathbin}{MnSyC}{"7C}
\DeclareMathSymbol{\diamonddot}{\mathbin}{MnSyC}{"7E}
\DeclareFontShape{U}{MnSymbolC}{m}{n}{
    <-6>  MnSymbolC5
   <6-7>  MnSymbolC6
   <7-8>  MnSymbolC7
   <8-9>  MnSymbolC8
   <9-10> MnSymbolC9
  <10-12> MnSymbolC10
  <12->   MnSymbolC12}{}

\begin{document}


\title{$\mOne$-Loop Theory} 



\author{Alexander S. Glasser}
\affiliation{
Princeton Plasma Physics Laboratory, Princeton University, Princeton, New Jersey 08543\\
Department of Astrophysical Sciences, Princeton University, Princeton, New Jersey 08544
}
\author{Hong Qin}
\affiliation{
Princeton Plasma Physics Laboratory, Princeton University, Princeton, New Jersey 08543\\
Department of Astrophysical Sciences, Princeton University, Princeton, New Jersey 08544
}


\date{\today}

\hyphenpenalty=1000
\begin{abstract}
A new formalism for lattice gauge theory is developed that preserves Poincar\'e symmetry in a discrete universe. We define the $\mOne$-loop, a generalization of the Wilson loop that reformulates classical differential equations of motion as identity-valued multiplicative loops of Lie group elements of the form ${[g_1\cdots g_n]=\mOne}$. A lattice Poincar\'e gauge theory of gravity is thus derived that employs a novel matter field construction and recovers Einstein's vacuum equations in the appropriate limit.
\end{abstract}
\hyphenpenalty=50


\maketitle 



\section{Introduction}

In recent decades, Lorentz invariance has been experimentally reaffirmed to increasingly high precision \cite{kostelecky_data_2011}. Nevertheless, the breakdown in the structure of space-time eventuated by many theories of quantum gravity \cite{garay_quantum_1995,magueijo_lorentz_2002,collins_lorentz_2004} continues to motivate scrutiny of this symmetry of nature \cite{beane_constraints_2014,lambiase_lorentz_2018,brun_detecting_2019}. Of particular present interest are those efforts to reconcile Lorentz invariance with a supposed discreteness of space-time. Some success in this direction has been achieved in loop quantum gravity \cite{rovelli_reconcile_2003} and causal set theory \cite{bombelli_discreteness_2009}. In a classical setting, Regge calculus \cite {regge_general_1961} has also been championed as a discrete, generally covariant theory of gravity \cite{feinberg_lattice_1984}, though its geometries' inequivalence under gauge transformations challenges that view \cite{loll_discrete_1998}.

In the present work, we develop a new formalism that reconciles Poincar\'e invariance with a discrete, but otherwise classical, universe. Our own motivation toward this effort arises from a desire to formulate algorithms that exactly conserve energy and momentum in simulations of classical physical systems. Because the space-times of algorithms are necessarily discrete, and the Noether symmetries \cite{noether_invariant_1971} they model necessarily continuous, vital conservation laws are generally broken in any first principles simulation. This forfeiture of space-time symmetry is a central challenge of computational physics, whose resolution bears upon questions of theoretical physics as well.

$\mOne$-loop theory is here introduced as a formalism for a lattice gauge theory of the Poincar\'e group, ${\mP=\mmT\rtimes SO^+(3,1)}$. We adopt an unconventional view of Poincar\'e symmetry  that identifies $\mP$ as a gauge group of foreground physical fields, rather than the symmetry group of a background space-time. We correspondingly regard the lattice of $\mOne$-loop theory as a mere graph, rather than an embedding in a continuous space-time, possessing dimensionality and extent.

The dynamical framework we adopt relinquishes Lagrangian and Hamiltonian formalism, and instead reformulates Yang-Mills \cite{yang_conservation_1954} equations of motion directly in a discrete, gauge-invariant construct we call the $\mOne$-loop. The $\mOne$-loop generalizes the Wilson loop \cite{wilson_confinement_1974} and derives its physics from a conserved current $\mJ$, rather than a Lagrangian $\mL$ or Hamiltonian $\mH$. $\mOne$-loop dynamics are described not by pointwise-defined differential equations---${\ws{E}(\mL)=0}$ or ${\d{}{t}=\{\cdot,\mH\}}$, say---but by finite lattice loops of Lie group elements whose composition evaluates to the identity: ${[g_1\cdots g_n](\mJ)=\mOne}$.

Briefly, the basic $\mOne$-loop for a gauge group $G$ is
\begin{eqn}
\delta\Omega\cdot J&=\mOne.
\label{OneLoopFieldEqn}
\end{eqn}
This relation recovers its Yang-Mills counterpart in the continuous space-time limit,
\begin{eqn}
\delta\mD A+j&=0.
\label{YangMillsFieldEqn}
\end{eqn}
The current $J$ and holonomy $\Omega$ of Eq.~(\ref{OneLoopFieldEqn}) are $G$-valued lattice loops, and ${\mOne\in G}$ denotes the identity. $\delta$ in Eq.~(\ref{OneLoopFieldEqn}) is a covariant codifferential redefined as a map between $G$-valued loops, satisfying ${\delta^2\Omega=\mOne}$ and ${(\delta\alpha)^{-1}=\delta(\alpha^{-1})}$ $\forall$ $\alpha$. Thus, ${\delta J=\mOne}$ follows from Eq.~(\ref{OneLoopFieldEqn}). This $\mOne$-loop conservation law forms a lattice counterpart to the Yang-Mills relation ${\delta j=0}$. Whereas Yang-Mills theories are defined for compact, reductive gauge groups, however, $\mOne$-loop theory is designed for the gauge groups of reductive Cartan geometries \cite{cartan_les_1926,sharpe_differential_1997}, such as $\mP$.

To construct a Poincar\'e $\mOne$-loop theory, a $\mP$-valued current $J$ will be defined. In fact, by requiring that currents transform in the adjoint representation, the $\mOne$-loop formalism uniquely determines $J$ from a mere choice of matter field. In this work, $J$ is thereby constructed from a recently defined Poincar\'e representation \cite{glasser_lifting_2019,glasser_restoring_2019}, the 5-vector $\Phi$. The $\mP$-valued holonomy $\Omega$ will likewise be formed from a Poincar\'e gauge field, $A$. Finally, leveraging ideas from Cartan geometry, we will define the operator $\delta$ in a manner comparable to the Wilson loop reconstruction of the covariant derivative $\mD$.

The resulting $\mOne$-loop theory constitutes a lattice Poincar\'e gauge theory of gravity. We will demonstrate that in the torsionless continuum limit, this theory recovers Einstein's vacuum equations and its fields evolve along geodesics. In the appropriate limit, therefore, Poincar\'e $\mOne$-loop theory accords with general relativity \cite{einstein_gr_1915} in vacuum. In the presence of matter, however, torsion and angular momentum play important dynamical roles in $\mOne$-loop theory---as they do in most Poincar\'e gauge theories of gravity \cite{kibble_lorentz_1961,sciama_physical_1964,hehl_spin_1973,hehl_spin_1974,hehl_general_1976,hehl_four_1980,trautman_fiber_1980,popov_theory_1975,popov_einstein_1976,tseytlin_poincare_1982,aldrovandi_complete_1984,aldrovandi_natural_1986,aldrovandi_quantization_1988}. We shall contextualize $\mOne$-loop theory within this existing literature. In its continuum limit, $\mOne$-loop theory will be seen to recover the field equations of a less-studied Poincar\'e gauge theory \cite{popov_theory_1975,popov_einstein_1976,tseytlin_poincare_1982,aldrovandi_complete_1984,aldrovandi_natural_1986,aldrovandi_quantization_1988}.

The remainder of this paper is organized as follows: section \ref{WilsonLoopSect} motivates the $\mOne$-loop as a natural generalization of the Wilson loop; section \ref{U1Warmup} further motivates $\mOne$-loop theory by studying a $U(1)$ lattice gauge theory; section \ref{Poincare1LoopTheory} introduces Poincar\'e $\mOne$-loop theory and comprises the core of this paper; section \ref{1LoopContinuumLimit} compares the continuum limit of $\mOne$-loop theory with existing gauge theories of gravity; and section \ref{conclusionSect} concludes.

\section{A Motivating Aside: Dynamical Variables for Gauge Theories\label{WilsonLoopSect}}

We briefly review an argument \cite{wu_concept_1975} for the naturalness of Wilson loops as dynamical variables for gauge fields. The experimentally-confirmed Aharonov-Bohm effect \cite{aharonov_significance_1959} demonstrates that the 2-form Faraday tensor ${F(x)\in\Lambda^2[M,\mathfrak{u}(1)]}$ under-describes the effects of the electromagnetic gauge field. On the other hand, the gauge field ${A(x)\in\Lambda^1[M,\mathfrak{u}(1)]}$ over-describes them: $A(x)$ can be freely gauge-transformed without physical consequence. Physical gauge-theoretic dynamical variables are therefore to be found somewhere `between' $A$ and $F$.

For abelian gauge theories, the group-valued Wilson loop \cite{weyl_elektron_1929}
\begin{eqn}
W_C=P\exp\left[i\oint_CA(x)\right]
\label{WilsonLoop}
\end{eqn}
satisfies both criteria of a dynamical variable; it is gauge-invariant and captures the Aharonov-Bohm effect. In non-abelian gauge theories, however, $W_C$ generally transforms nontrivially under a gauge transformation; although the path-ordering operator $P$ ensures the gauge invariance of $W_C$ at all intermediate points of the loop $C$, its basepoint $x_0$ leads $W_C$ to transform as the adjoint ${g(x_0)W_C g^{-1}(x_0)}$. As a result, the invariant dynamical variable in non-abelian gauge theories is defined by the trace of Eq.~(\ref{WilsonLoop}), that is, $\ws{Tr}[W_C]$.

In the present effort, we pursue a slightly different strategy in defining physical variables. In particular, we regard as physical variables only those group-valued loops that evaluate to the identity element, $\mOne$:
\begin{eqn}
\mOne_C=P\exp\left[i\oint_C(A|j)(x)\right]=\mOne.
\label{OneLoop}
\end{eqn}
In this expression, we have generalized the integrand of Eq.~(\ref{WilsonLoop}), allowing it to be either the gauge field $A(x)$ or the current $j(x)$, depending on the point ${x\in C}$. Like the gauge field, the current $j(x)$ of an arbitrary gauge theory is a $\mg$-valued 1-form \cite{bleecker_gauge_2005}. As such, $\mOne_C$  generalizes the Wilson loop to allow for its dependence on ${j(x)\in\Lambda^1[M,\mg]}$, while restricting it to be identity-valued. In what follows, we refer to a loop in the form of Eq.~(\ref{OneLoop}) as a $\mOne$-loop. When $\mOne$-loops are defined on a lattice, they serve as discrete, gauge-invariant counterparts to classical physics' pointwise-defined differential equations of motion.

Unlike $W_C$, the $\mOne$-loop is a suitable physical variable for an arbitrary gauge theory; because of its identity value, $\mOne_C$ is gauge-invariant for abelian and non-abelian groups alike. $\mOne_C$ may also be inverted or cyclically permuted without penalty; neither the orientation nor the basepoint $x_0$ of $C$ affects its evaluation to $\mOne$.

We emphasize two additional properties of $\mOne_C$, shared by $W_C$. First, whereas $F(x)$ is evaluated at a point in space-time, the spatio-temporal extent of the loop $C$ is non-vanishing. This motivates Wilson's exploration of gauge theories on a discrete lattice, where loops are necessarily non-vanishing and perhaps more readily defined. Second, $\mOne_C$ and $W_C$ both reveal the naturalness of working with Lie group elements rather than Lie algebra elements. For example, an observed phase difference in the Aharonov-Bohm experiment only fixes the integral in Eq.~(\ref{WilsonLoop}) up to integer multiples of ${2\pi}$. Its Lie algebra value is therefore unobservable and indeterminate, while its corresponding group element, by contrast, is fully specified.

\section{Attempting a $\mOne$-loop\\$U(1)$ Lattice Gauge Theory\label{U1Warmup}}

We first motivate $\mOne$-loop theory with a familiar gauge theory---scalar QED. We begin by recalling its classical equations of motion, derived from the Lagrangian ${\mL=-(\mD^\mu\phi)^*(\mD_\mu\phi)-m^2\phi^*\phi-\frac{1}{4}F^{\mu\nu}F_{\mu\nu}}$ in continuous ${\mR^{3,1}}$ space-time with flat metric ${g_{\mu\nu}(x)=\eta_{\mu\nu}}$ of signature ${(-\text{+++})}$:

\begin{subequations}
\begin{alignat}{1}
\mD^\mu F_{\mu\nu}+ej_\nu&=0
\label{U1ClassicalEOM_field}\\
(\mD^\mu\mD_\mu-m^2)\phi&=0,
\label{U1ClassicalEOM_matter}
\end{alignat}
\end{subequations}
where ${F_{\mu\nu}=\partial_\mu A_\nu-\partial_\nu A_\mu}$, ${\mD_\mu\phi=(\partial_\mu-ieA_\mu)\phi}$, and ${A_\mu\md x^\mu\in\Lambda^1[\mR^{3,1},\mathfrak{u}(1)]}$. Like the gauge field, the current ${j_\mu=i\left[(\mD_\mu\phi)^*\phi-\phi^*(\mD_\mu\phi)\right]}$ is a $\mathfrak{u}(1)$-valued 1-form, ${j_\mu\md x^\mu\in\Lambda^1[\mR^{3,1},\mathfrak{u}(1)]}$. For convenience, we recall that ${\mD^\mu F_{\mu\nu}=\partial^\mu F_{\mu\nu}}$ due to the trivial abelian adjoint action of ${g\in U(1)}$ on ${X\in\mathfrak{u}(1)}$, that is, ${\ws{Ad}_gX=gXg^{-1}=X}$.

Our strategy requires that all dynamical equations are re-expressed as $\mOne$-loops in the form of Eq.~(\ref{OneLoop}). To that end, we first restate the gauge field equation of motion by formally exponentiating the ${\mathfrak{u}(1)}$-valued Eq.~(\ref{U1ClassicalEOM_field}):
\begin{eqn}
\exp\big(\mD^\mu F_{\mu\nu}+ej_\nu\big)=\mOne.
\label{stepToGroupElements}
\end{eqn}
While identity-valued, as desired, this relation for the pointwise-defined Faraday tensor $F_{\mu\nu}$ must be further re-expressed as a loop integral of $A_\mu$ and $j_\mu$, as in Eq.~(\ref{OneLoop}).

We therefore reconstruct Eq.~(\ref{stepToGroupElements}) on a hypercubic lattice ${\{\v{n}\}=\mZ^4}$, letting Greek indices $\{\alpha,\beta,\dots\}$ correspond to lattice directions $\{t,x,y,z\}$. First, we derive a discrete form of Eq.~(\ref{U1ClassicalEOM_field}) from a lattice action $S$ defined over $\mZ^4$:
\begin{eqn}
S=\sum\limits_{\v{n}\in\raisebox{-.8pt}{\footnotesize$\mZ^4$}}\left[-\frac{1}{4}F^{\mu\nu}\n F_{\mu\nu}\n+ej_\mu\n A^\mu\n\right],
\label{DiscMaxwellAction}
\end{eqn}
where ${F_{\mu\nu}=\md_\mu^+A_\nu\n-\md_\nu^+A_\mu\n}$ with finite difference operator ${\md_\mu^\pm f\n=\pm(f[\v{n}\pm\hat{\mu}]-f\n)}$. For now, we regard ${j_\mu\n}$ as arbitrary and independent of $A^\nu$. Setting ${\partial S/\partial A^\nu\n=0}$, we derive the following discrete gauge field equation of motion from Eq.~(\ref{DiscMaxwellAction}):
\begin{eqn}
\eta^{\mu\sigma}\md^-_\mu F_{\sigma\nu}\n+ej_\nu\n=0.
\label{discreteGaugeFieldEOM}
\end{eqn}

\begin{figure}[b!]
\vspace*{-60pt}
\hbox{\hspace{-5pt}\includegraphics[width=88mm]{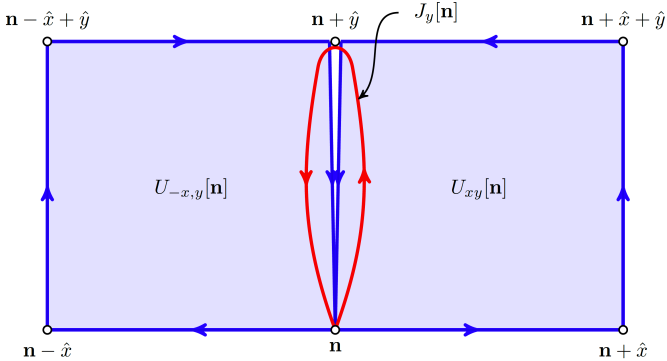}}
\setlength\abovecaptionskip{-35pt}
\setlength\belowcaptionskip{-5pt}
\caption{Noting the positive spatial signature of the metric $\eta_{\mu\nu}$, we depict ${G^x_{~y}\n=U_{xy}\n U_{-x,y}\n}$ and $J_y\n$. The current ${J_y\n}$ forms a round-trip loop along lattice edge ${[\v{n}|\v{n}+\hat{y}]}$. A $\mOne$-loop is thus formed by ${J_y\n G^t_{~y}\n G^x_{~y}\n G^z_{~y}\n=\mOne}$.}
\label{GxyPlot}
\end{figure}

We now substitute the left-hand side of Eq.~(\ref{discreteGaugeFieldEOM}) into the exponent of Eq.~(\ref{stepToGroupElements}) to discover the following $\mOne$-loop on $\mZ^4$:
\begin{eqn}
\mOne_\nu\n&=J_\nu\n\prod\limits_{\mu\neq\nu}G^\mu_{~\nu}\n\\
&=J_\nu\n\prod\limits_{\mu\neq\nu}\Big(U^\mu_{~~\nu}\n U^{-\mu}_{~~~~\nu}\n\Big)=\mOne.
\label{groupElementGaugeEOM}
\end{eqn}
This expression may be compared with Eq.~(\ref{OneLoop}). It is a $\mOne$-loop reformulation of Maxwell's equations in the desired form, ${[g_1\cdots g_n](j)=\mOne}$. In Eq.~(\ref{groupElementGaugeEOM}), $\nu$ is fixed and we have defined
\begin{eqn}
G^\mu_{~\nu}\n&=U^\mu_{~~\nu}\n U^{-\mu}_{~~~~\nu}\n\\
U^{\pm\mu}_{~~~~\nu}\n&=\exp(\eta^{\mu\mu}\log U_{\pm\mu,\nu}\n)\\
U_{\mu\nu}\n&=U_\nu\n^{-1}U_\mu[\v{n}+\hat{\nu}]^{-1}U_\nu[\v{n}+\hat{\mu}]U_\mu\n\\
U_\mu\n&=\exp\left(iA_\mu\n\right)\\
J_\mu\n&=\exp\left(iej_\mu\n\right).
\label{DefineLatticeVariables}
\end{eqn}
We have used the fact that $U(1)$ is abelian to freely factor the exponentiation of Eq.~(\ref{discreteGaugeFieldEOM}). For ${\mu=t}$, we note that ${U^\mu_{~~\nu}\n=U_{\mu\nu}\n^{-1}}$. A depiction of $J_y\n$ and $G^x_{~y}\n$ is rendered in Fig.~\ref{GxyPlot}.

Two obstructions to a $U(1)$ $\mOne$-loop theory are now seen plain. The first is, in part, aesthetic: The metric structure of Maxwell's equations is rather shoehorned into Eq.~(\ref{DefineLatticeVariables}). Beyond our desire to recover Eq.~(\ref{discreteGaugeFieldEOM}), there is no geometric motivation for the appearance of $\eta^{\mu\mu}$ in $U^{\pm\mu}_{~~~~\nu}\n$, nor is there a readily apparent generalization of Eq.~(\ref{groupElementGaugeEOM}) for a curved metric ${g_{\mu\nu}}$.

Second, although the $\mOne$-loop of Eq.~(\ref{groupElementGaugeEOM}) defines the desired dynamics for the $U(1)$ gauge field of lattice scalar QED, a complete theory must also specify dynamics for the matter field $\phi\n$ (and $j_\mu\n$ therewith). A variational Lagrangian approach can derive intuitive lattice discretizations of Eq.~(\ref{U1ClassicalEOM_matter}), such as \cite{shi_simulations_2018}
\begin{eqn}
\hspace{-5pt}\left[\mOne-S^{-\mu}e^{ieA_\mu\n}\right]_\mu&\left[e^{-ieA_\mu\n}S^\mu-\mOne\right]^\mu\phi\n-m^2\phi\n=0,
\label{discreteEOM}
\end{eqn}
 where ${S^{\pm\mu}\left(e^{A_\nu\n}f\n\right)=e^{A_\nu[\v{n}\pm\hat{\mu}]}f[\v{n}\pm\hat{\mu}]}$ defines the shift operator $S^{\pm\mu}$, and a Minkowski-signed sum over $\mu$ is implicit.
 
 However, while Eq.~(\ref{groupElementGaugeEOM}) formulates the gauge field equation of motion as a $\mOne$-loop, Eq.~(\ref{discreteEOM}) is not of this form. Indeed, because the matter field equation of motion---Eq.~(\ref{U1ClassicalEOM_matter})---is not Lie-algebra-valued, it offers no such re-expression. To fully define a  $\mOne$-loop theory, therefore, we must find an alternative group-valued representation of matter field dynamics.

A solution to this impasse is suggested by the following observation: The vanishing divergence of the energy-momentum tensor nearly enforces a system's equations of motion. This may be seen for Klein-Gordon and Dirac field theories as follows:
 \begin{eqn}
\partial^\mu T_{\mu\nu}^\phi&=\partial^\mu\left[\partial_\mu\phi\partial_\nu\phi+\eta_{\mu\nu}\mL^\phi\right]\\
&=\partial_\nu\phi\left[\partial^2\phi-m^2\phi\right]\\
\partial^\mu T_{\mu\nu}^\psi&=\partial^\mu\left[i\bar{\psi}\gamma_\mu\partial_\nu\psi-\eta_{\mu\nu}\mL^\psi\right]\\
&=\partial_\nu\bar{\psi}\left[-i\gamma^\mu\partial_\mu\psi+m\psi\right]+\left[i\partial_\mu\bar{\psi}\gamma^\mu+m\bar{\psi}\right]\partial_\nu\psi,
\label{consLawsAreEOM}
\end{eqn}
where we have defined ${2\mL^\phi=-\partial^\sigma\phi\partial_\sigma\phi-m^2\phi^2}$ and ${\mL^\psi=i\bar{\psi}\gamma^\sigma\partial_\sigma\psi-m\bar{\psi}\psi}$. We conclude from Eq.~(\ref{consLawsAreEOM}) that, wherever the matter fields $\phi$ and $\psi$ are non-constant, their respective equations of motion are enforced by energy-momentum conservation---that is, by ${\mmT}$ translation symmetry. (A similar result obtains in theories of fluids when their dynamical equations are written in conservative form.) In this sense, the energy-momentum tensor contains comparable information to a theory's Lagrangian or Hamiltonian.

Thus, for a gauge theory of a group ${G\supset\mmT}$, matter field dynamics may be determined by its $G$-valued conservation law. To address both aforementioned obstructions, therefore, we shall apply the preceding construction of ${U(1)}$ theory toward a $\mOne$-loop lattice gauge theory of the Poincar\'e group, ${\mP=\mmT\rtimes SO^+(3,1)}$. We will form a ten-component, Poincar\'e-valued energy-momentum $\mJ$ from a Poincar\'e representation---the 5-vector $\Phi$. In lieu of applying variational derivatives to derive Euler-Lagrange equations ${\ws{E}[\mL(\Phi)]=0}$, or a Poisson bracket to derive flows ${\dot{\Phi}=\{\Phi,\mH\}}$, we shall construct a $\mOne$-loop to discover the dynamics of $\Phi$ from $\mJ$. We thereby recover matter field dynamics from a ${\mP}$-valued $\mOne$-loop lattice gauge theory, whose gravitational dynamics will be explored in the continuum limit.

\section{$\mOne$-loop Poincar\'e Gauge Theory\label{Poincare1LoopTheory}}

We first define the Poincar\'e representation $\Phi$ on the hypercubic infinite lattice ${\{\v{n}\}=\mZ^4}$. We let Latin indices ${a,b,{\dots}\in\{t,x,y,z\}}$ correspond to lattice directions and let Greek indices ${\alpha,\beta,{\dots}\in\{0,1,2,3\}}$ denote internal Lorentz degrees of freedom. Following \cite{glasser_lifting_2019,glasser_restoring_2019}, we define the 5-vector $\Phi\n$ to be a 5-component vector at lattice vertex ${\v{n}\in\mZ^4}$ that gauge transforms under local Poincar\'e transformations, as follows:
\begin{eqn}
\Phi'\n&=g\n\tleft\Phi\n\\
&=(\Lambda,\vphi)\n\tleft\Phi\n\\[3pt]
&=\left[\begin{matrix}\Lambda^\mu_{~\nu}&\v{0}\\\vphi_\nu&1\end{matrix}\right]\left[\begin{matrix}\pi^\nu\\\phi\end{matrix}\right]\\[3pt]
&=\left[\begin{matrix}\Lambda^\mu_{~\nu}\pi^\nu\\\phi+\vphi_\nu\pi^\nu\end{matrix}\right].
\label{definePhiLatticeFieldAction}
\end{eqn}
Here, $'$ denotes a gauge transformation, $\tleft$ denotes a left group action, and $\n$ is often omitted for brevity. Thus, ${\Phi\n}$ is a real representation of the Poincar\'e group, ${\mP\rightarrow GL_5\big(\{\Phi\},\mR\big)}$, a transpose of the Bargmann representation \cite{bargmann_unitary_1954} of space-time transformations. The 5-vector formalism was previously adopted in the Duffin-Kemmer system \cite{rabin_homology_1982}, however, the gauge transformation of Eq.~(\ref{definePhiLatticeFieldAction}) has not been identified in such work.

We next describe a Poincar\'e gauge field along the lattice links of $\mZ^4$. We let ${\mmp=\ws{Lie}(\mP)}$, ${\mathfrak{h}=\ws{Lie}(H)}$ and ${\mt=\ws{Lie}(\mmT)}$ denote Lie algebras, where ${H=SO^+(3,1)}$. We define the lowered-index vierbein ${e_{\mu a}\n=\eta_{\mu\nu}e^\nu_{~a}\n}$ to be a $\mt$-valued translation gauge field on the link ${[\v{n}|\v{n}+\hat{a}]}$. We likewise define the $\mathfrak{h}$-valued gauge field ${\Gamma^\mu_{~\nu a}\n}$, which couples to ${e_{\mu a}\n}$ through the group action of $\mP$. To establish notation, we thereby construct a covariant derivative of the 5-vector as follows:
\begin{eqn}
\mD^+_a\Phi\n&=\Phi^+_a\n-\Phi\n\\
&=U_a\n^{-1}\cdot\Phi[\v{n}+\hat{a}]-\Phi\n\\
\left[\begin{matrix}\mD^+_a\pi^\mu\n\\[3pt]\mD^+_a\phi\n\end{matrix}\right]&=\left[\begin{matrix}(\pi^+_a)^\mu\n\\\phi^+_a\n\end{matrix}\right]-\left[\begin{matrix}\pi^\mu\n\\\phi\n\end{matrix}\right]\\
&=\exp\hspace{-2pt}\left[\begin{matrix}\Gamma^\mu_{~\nu a}\n & \v{0}\\e_{\nu a}\n & 0\end{matrix}\right]^{-1}\left[\begin{matrix}\pi^\nu\\\phi\end{matrix}\right][\v{n}+\hat{a}]-\left[\begin{matrix}\pi^\mu\\\phi\end{matrix}\right]\n.
\label{CovDerivDefine5Vector}
\end{eqn}
A backward covariant derivative is analogously defined:
\begin{eqn}
\mD^-_a\Phi\n&=\Phi\n-\Phi^-_a\n\\
&=\Phi\n-U_a[\v{n}-\hat{a}]\cdot\Phi[\v{n}-\hat{a}].
\end{eqn}
The parallel transport operator $U_a\n$ gauge transforms as usual: ${U_a'\n=g[\v{n}+\hat{a}]U_a\n g\n^{-1}}$, where ${g\in\mP}$.

The $\mOne$-loop dynamics of the 5-vector must derive from a Lie-algebra-valued current ${\mJ\in\Lambda^1[\mZ^4,\mmp]}$. Since $\mOne$-loop theory forgoes a Hamiltonian or Lagrangian structure, the properties of this current must be independently defined. We therefore pause to define $\mJ$ for a general lattice gauge theory in the $\mOne$-loop formalism.

\textbf{Definition:} Let ${G\rightarrow GL(V)}$ be a representation of a Lie group $G$ on a matter field ${\xi\n\in V}$ valued in vector space $V$. Let ${\{\latlink\}=V\times V\times GL(V)}$ denote the space of data determining a lattice edge in $\mZ^d$. We define the \emph{$\mOne$-loop current} ${\mJ\in\Lambda^1[\mZ^d,\mg]}$ to be a $G$-equivariant ${\text{$\mg$-valued}}$ 1-form, where ${\mg=\ws{Lie}(G)\subset\mathfrak{gl}(V)}$. In particular, ${\mJ:\{\latlink\}\rightarrow\mg}$ is required to satisfy ${\mJ\circ\Psi_g=\ws{Ad}_g\circ\mJ}$ ${\forall~g\in G}$, where $\Psi_g$ denotes the gauge transformation of a lattice edge, as follows:
\begin{eqn}
\begin{alignedat}{3}
{\Psi_g:}~~~\{\latlink\}~~&\rightarrow&~~&\{\latlink\}\\
(\xi_1,\xi_2,U)~~&\mapsto&&(g\tleft\xi_1,~h\tleft\xi_2,~h\cdot U\cdot g^{-1})
\label{Psi_g_defn}
\end{alignedat}
\end{eqn}
for ${h\in G}$ arbitrary.

Thus, an arbitrary gauge transformation on $\mZ^4$ maps ${\mJ_a\n}$ to ${\mJ_a'\n=\mJ(\Psi_g(\latlinkna))=g\n\mJ_a\n g\n^{-1}}$, where ${\mJ_a\n=\mJ(\latlinkna)}$ and ${\latlinkna=\big(\xi\n,\xi[\v{n}+\hat{a}],U_a\n\big)\in\{\latlink\}}$. The preceding definition of the current thereby enforces an adjoint action befitting the red loop depicted in Fig.~\ref{GxyPlot}.

This definition uniquely determines the $\mOne$-loop current of 5-vector lattice gauge theory. To see how, we first note that $\mJ_a\n$ can only depend on ${\big\{\Phi\n,\Phi[\v{n}+\hat{a}],U_a\n\big\}}$. Furthermore, since the $G$-equivariance of $\mJ_a\n$ precludes the appearance of ${g[\v{n}+\hat{a}]}$ in its gauge transformations, the dependence of ${\mJ_a\n}$ on ${\{\Phi[\v{n}+\hat{a}],U_a\n\}}$ is limited to the pairing ${\Phi^+_a\n=U_a\n^{-1}\Phi[\v{n}+\hat{a}]}$. (This construction accounts for the permissible arbitrariness of $h$ in the definition of $\Psi_g$ in Eq.~(\ref{Psi_g_defn}).) We must therefore solve for a $\mmp$-valued current ${\mJ_a\n=\mJ(\Phi,\Phi^+_a)}$ that transforms in the adjoint Poincar\'e representation---that is
\begin{eqn}
\ws{Ad}_{(\Lambda,\vphi)}\left[\begin{matrix}\Gamma & \v{0}\\e & 0\end{matrix}\right]&=\left[\begin{matrix}\Lambda & \v{0}\\\vphi & 1\end{matrix}\right]\left[\begin{matrix}\Gamma & \v{0}\\e & 0\end{matrix}\right]\left[\begin{matrix}\Lambda & \v{0}\\\vphi & 1\end{matrix}\right]^{-1}\\[3pt]
&=\left[\begin{matrix}\Lambda\Gamma\Lambda^{-1} & \v{0}\\(\vphi\Gamma+e)\Lambda^{-1} & 0\end{matrix}\right].
\label{adjointTransf}
\end{eqn}

Studying the Poincar\'e transformation of $\Phi$ in Eq.~(\ref{definePhiLatticeFieldAction}), up to a multiplicative constant there is found to be a unique such current:
\begin{eqn}
\mJ_a\n&=\left[\begin{array}{c|c}L^\mu_{~\nu a}\n & \v{0}\\[3pt]\hline T_{\nu a}\n & 0\end{array}\right]\\[3pt]
&=\frac{1}{2}\left[\begin{array}{c|c}\pi^\mu\boxtimes(\pi^+_a)_\nu & \v{0} \\[2pt]\hline\vphantom{\bigg|}\pi_\nu\phi^+_a-\phi(\pi^+_a)_\nu & 0\end{array}\right].
\label{defineTheCurrent}
\end{eqn}
Here, ${\pi_\nu=\pi^\sigma\eta_{\sigma\nu}}$ and ${\boxtimes:\mt^*\times\mt\rightarrow\mathfrak{h}}$ is the box map
\begin{eqn}
x^\mu\boxtimes y_\nu&=x\boxtimes(y^T\eta)=\Big[yx^T-xy^T\Big]\eta,
\label{boxproduct}
\end{eqn}
where $\eta$ denotes the ${4\times4}$ Minkowski matrix. ($\boxtimes$ roughly resembles the hat map of a 4-vector cross product.) It is readily confirmed that under the gauge transformation of Eq.~(\ref{definePhiLatticeFieldAction}), ${\mJ_a\n}$ transforms in the adjoint representation of Eq.~(\ref{adjointTransf}), that is, ${\mJ\circ\Psi_g=\ws{Ad}_g\circ\mJ}$ holds.

We note that ${\mD_a^+\Phi}$ can be freely substituted for ${\Phi^+_a}$ in Eq.~(\ref{defineTheCurrent}) without affecting its value. In particular, ${L^\mu_{~\nu a}=\frac{1}{2}\pi^\mu\boxtimes\mD_a^+\pi_\nu}$ since ${\pi^\mu\boxtimes\pi_\nu=0}$, and ${T_{\nu a}=\frac{1}{2}(\pi_\nu\mD^+_a\phi-\phi\mD_a^+\pi_\nu)}$ by a simple cancellation. In this form, $T_{\nu a}$ more closely resembles a current equivalent to the energy-momentum ${T^\phi_{\mu\nu}}$ of scalar field theory, as we now show.

We recall the two relations that define equivalence classes of nontrivial Noether currents in Lagrangian mechanics \cite{olver_applications_1986}. Two currents ${j_\mu[\phi]\cong\tilde{j}_\mu[\phi]}$ are equivalent (in the sense that their mutual conservation arises from the same Noether symmetry) if they
\begin{enumerate}[label=(\roman*)]
\item differ by an expression that vanishes on-shell; or
\item satisfy ${\partial^\mu(j_\mu[\phi]-\tilde{j}_\mu[\phi])=0}$ off-shell.
\end{enumerate}
We use both of these relations in the following calculation:
\begin{eqn}
T^\phi_{\mu\nu}&=\partial_\mu\phi\partial_\nu\phi-\frac{1}{2}\eta_{\mu\nu}\left(\partial^\sigma\phi\partial_\sigma\phi+m^2\phi^2\right)\\
&\stackrel{\mathclap{\normalfont\mbox{\tiny(i)}}}{\cong}\partial_\mu\phi\partial_\nu\phi-\frac{1}{2}\eta_{\mu\nu}\left(\partial^\sigma\phi\partial_\sigma\phi+\phi\partial^2\phi\right)\\
&\stackrel{\mathclap{\normalfont\mbox{\tiny(ii)}}}{\cong}\frac{1}{2}\Big[\partial_\mu\phi\partial_\nu\phi-\phi\partial_\mu\partial_\nu\phi\Big]\\
&\approx\frac{1}{2}e_\mu^{~a}\Big[\pi_\nu\partial_a\phi-\phi\partial_a\pi_\nu\Big]\approx e_\mu^{~a}T_{\nu a}|_{\mR^{3,1}}
\label{scalarTis5vectorT}
\end{eqn}
where we define ${T_{\nu a}|_{\mR^{3,1}}=\frac{1}{2}(\pi_\nu\partial_a\phi-\phi\partial_a\pi_\nu)}$, and set ${\partial_\mu\approx e_\mu^{~a}\partial_a}$ and ${\pi_\nu\approx\partial_\nu\phi}$ in the continuous, flat space-time limit, as in \cite{glasser_lifting_2019,glasser_restoring_2019}. Therefore, ${T_{\nu a}|_{\mR^{3,1}}}$ forms a continuous analogue of the 5-vector energy-momentum ${T_{\nu a}=\frac{1}{2}(\pi_\nu\mD^+_a\phi-\phi\mD_a^+\pi_\nu)}$.

We now take up our central effort: the construction of $\mP$-valued $\mOne$-loops that recover ${\delta\mD A+j=0}$ and ${\delta j=0}$ in the continuum limit. In Eq.~(\ref{defineTheCurrent}), we have already derived a group-valued current suitable for the matter field sector of such $\mOne$-loops. In particular, we define ${J:\{\latlink\}\rightarrow\mP}$ such that
\begin{eqn}
J(\latlinkna)=\exp(\kappa\mJ_a\n)
\end{eqn}
for some coupling constant $\kappa$. The first step in building the gauge field sector is similarly immediate: A realization of ${\mD A}$ suitable for $\mOne$-loop theory is found at once in the Wilson loop, a 2-form ${\Omega:\{\latsquare\}\rightarrow\mP}$ defined by ${\Omega(\latsquarenab)=U_{ab}\n}$, as specified in Eq.~(\ref{DefineLatticeVariables}). 

To complete this construction, we must define a $\mOne$-loop covariant codifferential operator $\delta$ and thereby assemble the desired $\mP$-valued analogue of the Yang-Mills field equation, ${\delta\Omega\cdot J=\mOne}$. We require $\delta$ to satisfy ${\delta^2\Omega=\mOne}$ and ${(\delta\alpha)^{-1}=\delta(\alpha^{-1})}$ for any form $\alpha$. The resulting $\mOne$-loop conservation law ${\delta J=\mOne}$ will then determine matter field dynamics, as anticipated in Eq.~(\ref{consLawsAreEOM}). Taken together, these relations ensure the integrability (or solvability) of $\mOne$-loop theory.

It is illuminating to first consider what a $\mOne$-loop conservation law ${\delta J=\mOne}$ could look like. We assume that it holds at each ${\v{n}\in\mZ^4}$, so that currents along all eight lattice links terminating on $\v{n}$ ought to play a role. Whereas ${\mJ_a\n}$ has endpoints at $\v{n}$, however, ${\mJ_a[\v{n}-\hat{a}]}$ has endpoints at ${\v{n}-\hat{a}}$. To incorporate the link ${[\v{n}-\hat{a}|\v{n}]}$, therefore, we substitute ${\Phi^+_a\rightarrow\Phi^-_a}$ in Eq.~(\ref{defineTheCurrent}), yielding the $\mmp$-valued current ${\mJ_{\bar{a}}\n}$, defined as follows:
\begin{eqn}
\mJ_{\bar{a}}\n=\mJ_a\n\Big|_{\Phi_a^+\rightarrow\Phi_a^-}=-\ws{Ad}_{U_a{[\v{n}-\hat{a}]}}\mJ_a[\v{n}-\hat{a}].
\label{transvertedCurrent}
\end{eqn}
The currents ${\mJ_a[\v{n}-\hat{a}]}$ and ${\mJ_{\bar{a}}\n}$ are defined by the same data, ${\latlinkan=\big(\Phi[\v{n}-\hat{a}],\Phi\n,U_a[\v{n}-\hat{a}]\big)}$. Their adjoint relationship in Eq.~(\ref{transvertedCurrent}), readily confirmed using $\Phi_a^-$ in Eq.~(\ref{defineTheCurrent}), demonstrates their compatibility under parallel transport. Eq.~(\ref{transvertedCurrent}) notably resembles the transformation of the Maurer-Cartan form $\omega_G$ under the inversion map $\iota$ \cite{sharpe_differential_1997}: ${(\iota^*\omega_G)(X_g)=-\ws{Ad}_g(\omega_G(X_g))}$ $\forall$ ${X_g\in T_gG}$.

\begin{figure}[h!]
\hbox{\hspace{-10pt}\includegraphics[width=95mm]{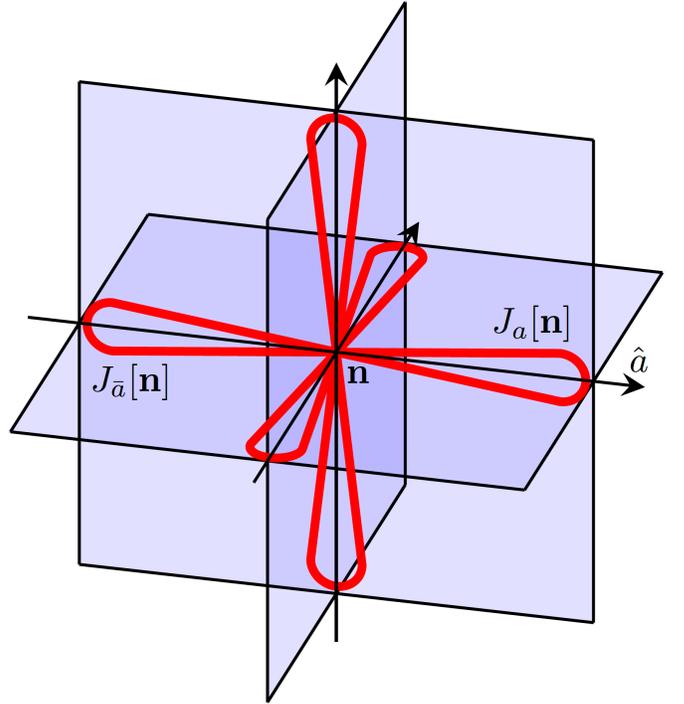}}
\caption{Six of the eight elements comprising the $\mOne$-loop conservation law ${\delta J\n=[g_1\cdots g_8](\mJ)=\mOne}$ are depicted at vertex $\v{n}$. In a continuous space-time limit, this conservation law recovers its Yang-Mills analogue, ${\delta j=0}$.}
\label{Current6Loops}
\end{figure}

An intuitive $\mOne$-loop analogue for ${\delta j=0}$, therefore, roughly takes the eight element form ${[g_1\cdots g_8](\mJ)=\mOne}$, as depicted in Fig.~\ref{Current6Loops}. The group elements $\{g_i\}$ must depend on $\mJ$ in a manner we shall make precise. To that end, we next revisit elements of Cartan geometry \cite{sharpe_differential_1997} and reinterpret them in the $\mOne$-loop formalism.

Let us recall ${\delta:\Lambda^\ell[M,\mR]\rightarrow\Lambda^{\ell-1}[M,\mR]}$, the codifferential of an $\ell$-form on a semi-Riemannian manifold $M$:
\begin{eqn}
\delta\alpha=g^{ab}\big[\mathrm{i}_{e_a}(\nabla_{e_b}\alpha)\big]&=\eta^{\mu\nu}\big[\mathrm{i}_{e_\mu}(\nabla_{e_\nu}\alpha)\big].
\label{codiffDef}
\end{eqn}
Here, $\{e_a\}$ is any local basis of ${TM}$ with inverse metric $g^{ab}$ and $\{e_\mu\}$ any local orthonormal basis. $\mathrm{i}_X$ denotes an interior product and $\nabla_X$ a covariant derivative with respect to ${X\in\mathfrak{X}(M)}$, where ${\mathfrak{X}(M)=\Gamma(TM)}$ denotes the set of vector fields on $M$. Eq.~(\ref{codiffDef}) is equivalent to the more widely used definition, ${\delta=\pm{\star}\md{\star}}$ \cite{eells_selected_1983}.

We now specialize $M$ to space-time, specifically, to the four-dimensional base space of a reductive Cartan geometry ${(P,\omega)}$. Here, ${\omega\in\Lambda^1[P,\mmp]}$ is a $\mmp$-valued Cartan connection on the right ${H}$-principal bundle ${P\xrightarrow{\pi}M}$, where ${\mmp=\mathfrak{h}\oplus\mt}$ is a direct sum of ${\ws{Ad}_H}$-modules and ${H=SO^+(3,1)}$. A Cartan connection establishes, by definition, an isomorphism ${\omega_p:T_pP\rightarrow\mmp}$ $\forall$ ${p\in P}$, such that $\omega^{-1}_p$ is everywhere well-defined. As such, the universal covariant derivative with respect to ${A\in\mmp}$ of any function ${f\in\mathcal{C}(P)}$ may be defined as follows \cite{sharpe_differential_1997}:
\begin{eqn}
\nabla_Af=[\omega^{-1}(A)]f.
\label{univCovDeriv}
\end{eqn}
${\nabla_A}$ differentiates $f$ with respect to the $\omega$-constant vector field ${\omega^{-1}(A)\in \mathfrak{X}(P)}$. In our notation, the universal $\nabla_A$ is distinguished from $\nabla_X$ merely by the distinct setting of ${A\in\mmp}$ and ${X\in\mathfrak{X}(M)}$.

An intuitive picture of this machinery will facilitate a $\mOne$-loop reformulation of $\delta$. First, we reinterpret the operator ${\nabla_{e_\mu}}$ in Eq.~(\ref{codiffDef}) as the universal covariant derivative of Eq.~(\ref{univCovDeriv})---in particular, we regard ${e_\mu\in\mt\subset\mmp}$ as a Lie algebra element. For any such ${e_\mu\in\mt}$, the $\omega$-constant vector field ${\omega^{-1}(e_\mu)}$ generates, by definition, geodesics on $P$. ${\mathrm{i}_{e_\mu}(\nabla_{e_\mu}\alpha)}$ therefore represents the change in $\alpha$ along the geodesic generated by ${\omega^{-1}(e_\mu)}$.

In this way, $\delta\alpha$ sums the change in $\alpha$ over orthonormal geodesics, weighted by a metric factor as in Eq.~(\ref{codiffDef}). In the $\mOne$-loop formalism, this metric structure is not provided by the base manifold (or lattice) but by the Lie algebra $\mmp$. In particular, we introduce the following nondegenerate, ${\ws{Ad}_H}$-invariant metric ${\langle \cdot,\cdot\rangle_\mmp}$ on $\mmp$:
\begin{eqn}
\langle A,B\rangle_\mmp&=\ws{Tr}\left(A\ms{\eta}B^T\ms{\eta}\right)\hspace{24pt}\text{where }\ms{\eta}=\left[\begin{matrix}\eta & \v{0}\\ \v{0} & 1\end{matrix}\right]\in\mR^{5\times5}\\
&=\ws{Tr}\left(\Gamma_A\eta\Gamma_B^T\eta\right)+e_A\eta e_B^T
\label{PoincareMetric}
\end{eqn}
$\forall$ ${A,B\in\mmp}$. Here, ${A=(\Gamma_A,e_A)\in\mmp}$ denotes a matrix Lie algebra element as appears in Eqs.~(\ref{CovDerivDefine5Vector}) and (\ref{adjointTransf}), with $e_A$ a row vector. $e_A\eta e_B^T$ may be recognized as the semi-Riemannian metric.

We now construct the $\mOne$-loop operator $\delta$. We first define a discrete Cartan connection ${\omega:\{\latlink\}\rightarrow\mmp}$ such that ${\exp(\omega_a\n)=U_a\n\in\mP}$. Since $\delta$ aggregates over orthonormal geodesics, we choose an arbitrary basis $\{e_\mu\}$ of ${\mt\subset\mmp}$ that is orthonormal with respect to ${\langle\cdot,\cdot\rangle_\mmp}$. We further define the neighborhood $\mathcal{N}_\v{n}$ of $\v{n}$, comprised of $\v{n}$ and its eight nearest neighbors in ${\mZ^4}$. Now, $\mathcal{N}_\v{n}$ is said to be \emph{rectified} if each connection ${\{\omega_a\n,\omega_{\bar{a}}\n\}}$ evaluates to a distinct basis vector in ${\{\pm e_\mu\}}$, e.g. ${\omega_x\n=-\omega_{\bar{x}}\n=e_\mu}$. To rectify $\mathcal{N}_\v{n}$, we apply suitable gauge transformations at each neighbor of $\v{n}$. We require, however, that the chosen transformations ${\{g[\v{n}\pm\hat{a}]\in\mP\}}$ preserve the `isomorphism' of $\omega$ on $\mathcal{N}_\v{n}$. In effect, the vierbeins ${e^\mu_{~a}\n}$ and ${e^\mu_{~\bar{a}}\n}$ should remain nonsingular as they are smoothly rectified toward ($\pm$) the identity matrix.

Concretely, this procedure yields transformed comparators of the form ${U_a'\n=g[\v{n}+\hat{a}]U_a\n=\exp(e_\mu)}$, where ${g\n=\mOne}$. Any matter or gauge field data that has already been defined within or adjoining $\mathcal{N}_\v{n}$ must also be gauge-transformed accordingly; therefore, not all lattice neighborhoods can be rectified simultaneously. We note that data as yet undefined need not (and of course cannot) be transformed in this way. We shall denote a rectified neighborhood by ${\bar{\mathcal{N}}_\v{n}}$. Crucially, we observe that $\bar{\mathcal{N}}_\v{n}$ defines a bijection, ${r:\{\pm\mu\}\rightarrow\{a,\bar{a}\}}$.

At last, we define ${\delta:\Lambda^\ell[\bar{\mathcal{N}}_\v{n},\mP]\rightarrow\Lambda^{\ell-1}[\bar{\mathcal{N}}_\v{n},\mP]}$ as a map between rectified $\mP$-valued forms. A rectified form ${\alpha\in\Lambda^\ell[\bar{\mathcal{N}}_\v{n},\mP]}$ is a form defined on a rectified neighborhood that transforms under a gauge transformation as ${\alpha'\n=g\n\alpha\n g\n^{-1}}$. (Intuitively, a rectified form is a closed loop with endpoints at $\v{n}$.) Clearly, $J$ and $\Omega$ are rectified forms on $\bar{\mathcal{N}}_\v{n}$, while ${U_a=\exp(\omega_a)}$ is not. $\delta$ is now readily defined:
\begin{eqn}
\delta J\n&=\exp\sum\limits_{\nu\in\{\pm\mu\}}\log J^\nu\n\\
\delta\Omega_b\n&=\exp\sum\limits_{\nu\in\{\pm\mu\}}\log\Omega^\nu_{~b}\n.
\label{deltaOneLoops}
\end{eqn}

\begin{figure}[b!]
\vspace*{-40pt}
\hbox{\includegraphics[width=89mm]{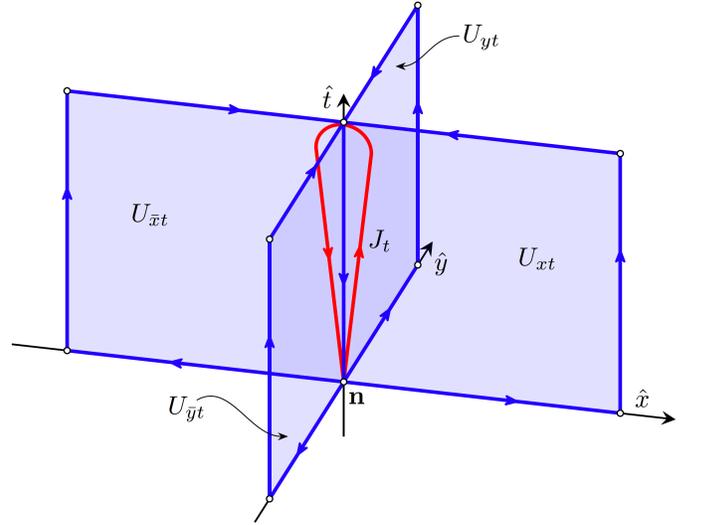}}
\setlength\abovecaptionskip{-15pt}
\setlength\belowcaptionskip{-5pt}
\caption{Four of the six holonomies comprising the $\mOne$-loop ${\delta\Omega_t\n J_t\n=\mOne}$ are depicted with $J_t\n$. ($U_{zt}$ and $U_{\bar{z}t}$ are not shown.) $\delta\Omega_t\n$ can be regarded as a `simultaneous' multiplication of the six holonomies adjoining lattice edge ${[\v{n}|\v{n}+\hat{t}]}$.}
\label{OneLoopPlot}
\end{figure}

The notation of Eq.~(\ref{deltaOneLoops}) requires some clarification. $\exp$ and $\log$ denote matrix exponentials and logarithms, respectively. As in Eq.~(\ref{DefineLatticeVariables}), a raised index indicates a metric factor, such that ${\log J^\mu=\eta^{\mu\mu}\log J_\mu}$. (This metric factor is now seen to arise from its associated $e_\mu$ geodesic.) Furthermore, $J_{\mu}\n$ denotes ${J_{r(\mu)}\n}$, the current on a single link in $\bar{\mathcal{N}}_\v{n}$, as determined by the bijection $r$. Similarly, ${\Omega_{\mu b}\n}$ denotes the holonomy ${\Omega_{r(\mu)b}\n}$ on a single plaquette. As seen in Eq.~(\ref{transvertedCurrent}), the `backward' currents ${\{\mJ_{\bar{a}}\}}$ are implicitly negated, as are the `reversed' holonomies $\{\log\Omega_{\bar{a}b}\}$. Thus, the intuitive picture of ${\delta J}$ (or ${\delta\Omega}$) as the metric-weighted change in $J$ (or $\Omega$) along geodesics is realized. 

The use of the logarithm in the definition of $\delta$ enables the `simultaneous' multiplication of non-abelian group elements---without preferential ordering. $\delta$ of a rectified 1-form (e.g. $\delta J$) can therefore be imagined as a simultaneous contraction of loops along all edges adjoining a vertex, and $\delta$ of a rectified 2-form (e.g. $\delta\Omega$) as the simultaneous contraction of loops along all plaquettes adjoining an edge. (See Figs.~\ref{Current6Loops} and \ref{OneLoopPlot}.)  The absence of ordering in these multiplications ensures that ${\delta^2\Omega=\mOne}$ and ${(\delta\alpha)^{-1}=\delta(\alpha^{-1})}$, properties of $\delta$ that are readily verified with Eq.~(\ref{deltaOneLoops}).

The basic $\mOne$-loop of Eq.~(\ref{OneLoopFieldEqn}) is now completely defined. Lattice fields are thus evolved by solving the $\mOne$-loops ${\delta\Omega_a\cdot  J_a=\mOne}$ and ${\delta J=\mOne}$ for their unknown data. We assume that this evolution proceeds time-slice by time-slice, and is therefore realizable by the following iterative algorithm:

\begin{enumerate}[label=(\roman*)]
\item Self-consistently initialize ${\Phi\n}$, ${\Phi[\v{n}+\hat{t}\hspace{1pt}]}$, ${\omega_a\n}$ and ${\omega_b[\v{n}+\hat{t}\hspace{1pt}]}$ $\forall$ ${\v{n}\in\{n_t=0\}}$, ${a\in\{t,\v{x}\}}$ and ${b\in\{\v{x}\}}$.
\item Since no $\mOne$-loop is completed by defining a gauge field along a temporal link ${[n_t=1|n_t=2]}$, any such link may be freely specified. Thus, assign the temporal gauge: ${\omega_t\n=e_0}$ $\forall$ ${\v{n}\in\{n_t=1\}}$.
\item Solve ${\delta J\n=\mOne}$ for $\Phi[\v{n}+\hat{t}\hspace{1pt}]$ $\forall$ ${\v{n}\in\{n_t=1\}}$.
\item Solve ${\delta\Omega_b\n J_b\n=\mOne}$ for ${\omega_b[\v{n}+\hat{t}\hspace{1pt}]}$ $\forall$ ${\v{n}\in\{n_t=1\}}$ and ${b\in\{\v{x}\}}$.
\item Return to (ii), assigning ${\omega_t\n=e_0}$ $\forall$ ${\v{n}\in\{n_t=2\}}$.
\end{enumerate}

A conceptually straightforward approach to calculating $\delta$ in steps (iii) and (iv) is to rectify a maximal set of disjoint neighborhoods ${\bar{\mathcal{M}}=\sqcup_\v{n}\{\bar{\mathcal{N}}_\v{n}\}}$ on ${\mZ^4|_{n_t=1}}$, solve for rectified forms on $\bar{\mathcal{M}}$, and then repeat for the as-yet-unrectified neighborhoods on ${\mZ^4|_{n_t=1}\backslash\bar{\mathcal{M}}}$. We note, however, that rectified forms on a neighborhood ${\mathcal{N}_\v{n}}$ are in fact invariant under the `$\v{n}$-adjacent' transformations ${\{g[\v{n}\pm\hat{a}]\}}$ (since their loops terminate at $\v{n}$). In principle, therefore, a more streamlined approach could solve all neighborhoods of $\mZ^4$ in parallel as if they were rectified, and afterward resolve any mismatches of gauge. We observe that the preceding algorithm maximally leverages the gauge-invariance of the $\mOne$-loop; the rectification of ${\mathcal{N}_\v{n}}$ by $\delta$ is made permissible by the gauge invariance of ${\delta J=\mOne}$ and ${\delta\Omega_b\n J_b\n=\mOne}$.

Let us consider the solvability of this algorithm. A self-consistent initialization of step (i) requires that ${\delta\Omega_t\n J_t\n=\mOne}$ $\forall$ ${\v{n}\in\{n_t=0\}}$. By specifying ${J_t\n\in\mP}$ first, and then the matter fields ${\Phi\n}$ and ${\Phi^+_t\n}$ comprising it, various suitable initial conditions are readily found. All steps of the algorithm are then immediately solvable, except perhaps step (iii). We note, however, that Eq.~(\ref{defineTheCurrent}) is linear in ${\Phi^+_a}$. Therefore, since every leg of ${\delta J\n}$ shares the same $\Phi\n$, a solution ${\Phi^+_t\n}$ to ${\delta J\n=\mOne}$ must exist. For completeness, we nevertheless note the following conditions on ${\Phi=[\pi^\mu;\phi]}$ necessary for the existence of a solution ${\Phi^+_t}$ to ${\mJ(\Phi,\Phi^+_t)=({\Gamma}^\mu_{~\nu},{e}_\nu)}$:
\begin{eqn}
\pi^\mu\boxtimes{e}_\nu+\phi{\Gamma}^\mu_{~\nu}&=0\\
\pi^\sigma{\Gamma}^{\tau\mu}+\pi^\tau{\Gamma}^{\mu\sigma}+\pi^\mu{\Gamma}^{\sigma\tau}&=0,
\label{consConditions}
\end{eqn}
where ${({\Gamma}^\mu_{~\nu},{e}_\nu)\in\mmp}$ and ${{\Gamma}^{\sigma\tau}={\Gamma}^\sigma_{~\nu}\eta^{\nu\tau}}$.

Furthermore, when the representation $\Phi$ is fully specified, the preceding algorithm is not only solvable but uniquely determined---up to arbitrary choices of gauge. In particular, once the mass and `time direction' of $\Phi$ are fixed (such that ${\pi^\mu\pi_\mu+m^2=0}$ and ${\pi^0>0}$ $\forall$ $\v{n}$, for example), the conservation law ${\delta J=\mOne}$ fully determines its evolution. Likewise, ${\delta\Omega\cdot J=\mOne}$ uniquely determines $\omega_a$.

We shall leave a more robust examination of the dynamics of $\Phi$ to future work. For now, having described an algorithm for the evolution of Poincar\'e $\mOne$-loop theory, we examine its physics in the continuum limit.





\section{Gravity in the $\mOne$-loop Formalism\label{1LoopContinuumLimit}}
We consider the continuum limit of $\mOne$-loop Poincar\'e lattice gauge theory. We denote our gauge field by ${\omega_a\n=A_a\n\in\mmp}$ and define its comparators with a lattice parameter $\Delta$, that is: ${U_a\n=\exp(\Delta A_a\n)}$.  Applying the BCH formula---see \cite{kogut_introduction_1979} Eq.~(8.7)---and expanding gauge fields at lattice points away from $\v{n}$---e.g. ${A_b[\v{n}\pm\hat{a}]=[A_b\pm\Delta\partial_aA_b+\frac{\Delta^2}{2}\partial_a^2A_b+\mO(\Delta^3)]}$---we find, in the ${\Delta\rightarrow0}$ limit:
\begin{eqn}
\log U_{ab}\n&=\Delta^2F_{ab}+\mO(\Delta^3)\\
\log U_{\bar{a}b}\n+\log U_{ab}\n&=\Delta^3\mD_aF_{ab}+\mO(\Delta^4),
\label{DeltaExpansionsOfUab}
\end{eqn}
where ${F_{ab}=\partial_{[a}A_{b]}-[A_a,A_b]}$ and ${\mD_a=\partial_a-[A_a,\cdot]}$. (Note, no index summation is implied in Eq.~(\ref{DeltaExpansionsOfUab}); we omit the conventional factor of $\frac{1}{2}$ in our notation for the antisymmetrization of indices; and the sign conventions in ${F_{ab}}$ and $\mD_a$ arise because the gauge field has a left action, $\tleft$.) Computing the Lie brackets in the definitions of $F_{ab}$ and $\mD_a$, we explicitly evaluate the fields of Eq.~(\ref{DeltaExpansionsOfUab}) for our $\mmp$-valued connection as follows:
\begin{subequations}
\begin{alignat}{1}
\hspace{-5pt}A_a&=\left[\begin{array}{c|c}\Gamma^\mu_{~\nu a} & \v{0}\\\hline e_{\nu a} & 0\end{array}
\right]\label{PoincareA}\\
\hspace{-5pt}F_{ab}&=\left[\begin{array}{c|c}F^\mu_{~\nu ab} & \v{0}\\\hline F_{\nu ab} & 0\end{array}\right]=\left[\begin{array}{c|c}\partial_{[a|}\Gamma^\mu_{~\nu |b]}-\Gamma^\mu_{~\sigma[a|}\Gamma^\sigma_{~\nu|b]} & \v{0}\\\hline\partial_{[a|}e_{\nu|b]}-e_{\sigma[a|}\Gamma^\sigma_{~\nu|b]} & 0\end{array}\right]\label{PoincareF}\\
\hspace{-5pt}\mD_cF_{ab}&=\left[\begin{array}{c|c}\partial_cF^\mu_{~\nu ab}-\Gamma^\mu_{~\sigma c}F^\sigma_{~\nu ab}+F^\mu_{~\sigma ab}\Gamma^\sigma_{~\nu c} & \v{0}\\\hline\partial_cF_{\nu ab}-e_{\sigma c}F^\sigma_{~\nu ab}+F_{\sigma ab}\Gamma^\sigma_{~\nu c} & 0\end{array}\right].
\label{PoincareDF}
\end{alignat}
\end{subequations}

We now substitute Eqs.~(\ref{deltaOneLoops}), (\ref{DeltaExpansionsOfUab}) and (\ref{PoincareDF}) into the $\mOne$-loop ${\delta\Omega_b\n J_b\n=\mOne}$ of Eq.~(\ref{OneLoopFieldEqn}), keeping terms to least order in $\Delta$. Working on $\bar{\mathcal{N}}_\v{n}$, we thus discover
\begin{eqn}
\mD^aF_{ab}+\kappa\mJ_b=\left[\begin{array}{c|c}\partial^aF^\mu_{~\nu ab}+\kappa L^\mu_{~\nu b} & \v{0}\\\hline\partial^aF_{\nu ab}-e_\sigma^{~a}F^\sigma_{~\nu ab}+\kappa T_{\nu b} & 0\end{array}\right]&=\mZero,
\label{continuum1Loop5VT}
\end{eqn}
where we have set ${\Gamma^\mu_{~\nu a}\n=0}$ and ${g^{ab}\n=\eta^{ab}}$. (In the continuum limit, rectification resembles a local application of Riemann normal coordinates.) ${L^\mu_{~\nu b}}$ and $T_{\nu b}$ in Eq.~(\ref{continuum1Loop5VT}) are assumed to be in the ${\Delta\rightarrow0}$ limit. Restoring $\Gamma^\mu_{~\nu a}$, we may re-express the field equations of Eq.~(\ref{continuum1Loop5VT}) more schematically as
\begin{eqn}
\partial R-[\Gamma,R]+\kappa L&=0\\
\partial S-[e,R]-[\Gamma,S]+\kappa T&=0
\label{schematicFieldEq}
\end{eqn}
where $R$ (i.e. $F^\mu_{~\nu ab}$) and $S$ (i.e. $F_{\nu ab}$) roughly represent space-time curvature and torsion, respectively---an interpretation we shall justify in Eq.~(\ref{FinGRLimit}). We emphasize that these field equations comprise the continuous limit of the well-posed discrete algorithm of the previous section.

Let us compare this result with existing gauge theories of gravity. The earliest attempt at a modern gauge theory of gravity was made in 1955 by Utiyama \cite{utiyama_invariant_1956}, who identified the Lorentz group as the relevant gauge group. ${SO^+(3,1)}$ is an instinctive fit for gravity, not least because the Lorentz field strength $F^\mu_{~\nu ab}$ essentially reproduces the Riemann tensor of curved space-time, as in  Eq.~(\ref{FinGRLimit}). Utiyama's formalism appears to suggest \cite{hammond_torsion_2002}, however, that the sole Noether current associated with gravity is angular momentum ($L$)---a result that perhaps underrates energy-momentum ($T$) as a source of gravitation.

Subsequent efforts were made to incorporate a more complete description of the gauge symmetries and conserved quantities of gravity. An examination of the literature reveals that, since the 1960s, at least two parallel tracks developed in Poincar\'e gauge theories of gravity. These might be called the L (Lagrangian) track \cite{kibble_lorentz_1961,sciama_physical_1964,hehl_spin_1973,hehl_spin_1974,hehl_general_1976,hehl_four_1980,trautman_fiber_1980} and the YM (Yang-Mills) track \cite{popov_theory_1975,popov_einstein_1976,tseytlin_poincare_1982,aldrovandi_complete_1984,aldrovandi_natural_1986,aldrovandi_quantization_1988}.

The widely studied L-track originated in the 1960s. Kibble \cite{kibble_lorentz_1961} and Sciama \cite{sciama_physical_1964} extended Utiyama's  gauge theory to the Poincar\'e group, yielding Einstein-Cartan-Sciama-Kibble gravity, or $U_4$ theory \cite{hehl_general_1976}. While the Poincar\'e gauge field curvatures of $U_4$ theory are identical to those of Eq.~(\ref{PoincareF}), the matter couplings of its field equations differ considerably from those of Eq.~(\ref{schematicFieldEq}). In its simplest form, $U_4$ theory couples angular momentum ($L$) not with curvature ($R$), but with a non-propagating torsion ($S$). Somewhat unexpectedly, therefore, angular momentum is coupled in $U_4$ theory to the gauge field curvature associated with the translation subgroup ${\mmT\subset\mP}$. The L-track hews to a Lagrangian formalism and, in all of its manifestations, derives from the \emph{terra firma} of a variational principle.

The YM-track originated in the 1970s with an attempt by Popov and Daikhin \cite{popov_theory_1975,popov_einstein_1976} to derive a more orthodox gauge theory of gravitation. This branch of Poincar\'e gauge theory is of particular relevance here, because its field equations \cite{tseytlin_poincare_1982,aldrovandi_complete_1984} are precisely recovered in the continuum limit of Poincar\'e $\mOne$-loop theory, as derived in Eqs.~(\ref{continuum1Loop5VT})-(\ref{schematicFieldEq}). Unlike those of the L-track, these field equations couple a propagating torsion ($\partial S$) to energy-momentum ($T$). The YM-track has proven to be underivable from a Lagrangian formalism \cite{tseytlin_poincare_1982,aldrovandi_complete_1984,aldrovandi_natural_1986,aldrovandi_quantization_1988}. It is perhaps unsurprising, then, that a new dynamical formalism such as $\mOne$-loop theory might, in its continuum limit,  rediscover it. We shall further characterize key results of the YM-track in our concluding discussion. For now, having contextualized the continuum limit of Poincar\'e $\mOne$-loop theory, we proceed to demonstrate its recovery of Einstein's vacuum equations.

We take a general relativistic (GR) limit of Eq.~(\ref{continuum1Loop5VT}) by imposing two additional assumptions upon it, namely, metric compatibility and zero torsion. The former---${\mD_cg_{ab}=0}$---may be established by defining a vanishing covariant derivative of the translation gauge field \cite{blagojevic_gravitation_2002}\cmmnt{Eq.~(3.43)}:
\begin{eqn}
0&=\mD_ae_{\mu b}\\
&=\partial_ae_{\mu b}+\Gamma^\sigma_{~\mu a}e_{\sigma b}+\Gamma^c_{~ba}e_{\mu c}.
\label{CovDivE}
\end{eqn}
Here, we have introduced the affine connection $\Gamma^c_{~ba}$, whose degrees of freedom are not independent and are fixed in terms of the Poincar\'e gauge fields by Eq.~(\ref{CovDivE}). The Riemann tensor is then defined as usual in terms of this affine connection:
\begin{eqn}
R^c_{~dab}=\partial_{[a|}\Gamma^c_{~d|b]}-\Gamma^c_{~e[a|}\Gamma^e_{~d|b]}.
\end{eqn}
The latter assumption, zero torsion, is defined as follows:
\begin{eqn}
S^c_{~ab}=\Gamma^c_{~[ab]}=0.
\label{ZeroTorsion}
\end{eqn}

We now substitute Eqs.~(\ref{CovDivE})-(\ref{ZeroTorsion}) to eliminate the Lorentz gauge field $\Gamma^\mu_{~\nu a}$ in Eq.~(\ref{PoincareF}). Simplifying, we find that in the GR limit
\begin{eqn}
F^\mu_{~\nu ab}&=e^{\mu d}e_{\nu c}R^c_{~dba}\\
F_{\nu ab}&=e_{\nu c}S^c_{~ab}=0,
\label{FinGRLimit}
\end{eqn}
where ${e^\mu_{~b}e_\mu^{~a}=\delta^a_b}$ and ${g_{ab}=e^\mu_{~a}\eta_{\mu\nu}e^\nu_{~b}}$. Therefore, the Riemann and torsion tensors are closely related to the gauge field curvatures defined in Eq.~(\ref{PoincareF}), as desired. Since ${F^{\mu\nu}_{~~ab}}$ is antisymmetric in its first two and last two indices, it further follows from Eq.~(\ref{FinGRLimit}) that, in the GR limit, $R^{cd}_{~~ab}$ is as well.

Finally, substituting Eq.~(\ref{FinGRLimit}) into the translation components of Eq.~(\ref{continuum1Loop5VT}) and setting ${T_{\nu b}=0}$, we thus recover Einstein's vacuum equations
\begin{eqn}
e_{\nu c}R^{ca}_{~~ba}=0,
\label{EinsteinVacEqn}
\end{eqn}
as desired.

\section{Discussion and Conclusions\label{conclusionSect}}

The $\mOne$-loop formalism has been demonstrated to successfully define a lattice gauge theory of the Poincar\'e group. By reinterpreting $\mP$ as an internal gauge group, $\mOne$-loop theory preserves Poincar\'e symmetry on a discrete lattice, and recovers general relativity in its torsionless, continuum vacuum limit. This new formalism comprises several technical innovations:

\begin{enumerate}[label=(\roman*)]
\item the $\mOne$-loop of Eq.~(\ref{OneLoop})---a relative of the Wilson loop that reconstitutes differential equations of motion;
\item the 5-vector $\Phi$ of Eq.~(\ref{definePhiLatticeFieldAction})---a new representation of the Poincar\'e group;
\item the definition of $\mOne$-loop current, which uniquely determines the Poincar\'e current ${\mJ_a}$ of Eq.~(\ref{defineTheCurrent}); and
\item the lattice covariant codifferential $\delta$ of Eq.~(\ref{deltaOneLoops}), motivated by Cartan geometry.
\end{enumerate}

The dynamics of the resulting Poincar\'e gauge theory are determined by the basic $\mOne$-loop ${\delta\Omega\cdot J=\mOne}$, as defined in Eq.~(\ref{OneLoopFieldEqn}). This $\mP$-valued analogue of a Yang-Mills field equation defines not only the dynamics of the Poincar\'e gauge field, but matter field dynamics as well. Indeed, matter field equations of motion are superfluous in $\mOne$-loop theory, as they follow from the  conservation of the $\mP$-valued Noether current, ${\delta J=\mOne}$, guaranteed in turn by ${\delta^2\Omega=\mOne}$. The $\mOne$-loop formalism thereby defines a computable, exactly-energy-momentum-conserving algorithm for the dynamics of a 5-vector matter field evolving under gravity.

A $\mOne$-loop theory is decidedly rigid in the sense that very few arbitrary choices are made in its construction. Given a $G$-representation and a reductive Cartan geometry with $\mg$-valued connection $\omega$ and base-space $M$, a corresponding $\mOne$-loop theory is already quite fixed: the hypercubic lattice $\mZ^d$ is defined such that ${d=\dim[M]}$; the holonomy $\Omega$ is fixed by $\omega$; the $\mOne$-loop current $J$ is fixed by the $G$-representation, as demonstrated for $G=\mP$ in Eq.~(\ref{defineTheCurrent}); and the interaction of matter and gauge fields is wholly determined by the $\mOne$-loop ${\delta\Omega\cdot J=\mOne}$. The definition of the operator $\delta$ is itself quite constrained by its need to satisfy ${\delta^2\Omega=\mOne}$.

The choice to relinquish a Lagrangian in favor of the $\mOne$-loop formalism was not undertaken without considerable effort by the authors to construct a satisfactory Lagrangian Poincar\'e lattice gauge theory. In a Lagrangian approach, Poincar\'e symmetry generators naturally arise as vector fields on space-time, which are ill-defined on a discrete lattice. An effort to `lift' these generators to vertical gauge symmetries of a discrete Lagrangian apparently requires the introduction of new fields that do not have a clear physical interpretation \cite{glasser_lifting_2019,glasser_restoring_2019}. More abstractly, this earlier work revealed a natural tension between the additive structure of a Lagrangian action---integrated over space-time or summed over lattice vertices---and the multiplicative group structure of the Poincar\'e symmetries.

A Hamiltonian approach was also considered, however, operator-based Hamiltonian theories are predicated on the evolution of a continuous time parameter that is unsuitable for computation. Although gauge-compatible splitting methods \cite{glasser_geometric_2020} enable the preservation of gauge structure in discrete-time Hamiltonian algorithms, it is unclear how such a splitting in time can be extended to a `four-dimensional splitting' over a space-time lattice.

$\mOne$-loop formalism was developed to address these challenges. It assumes a multiplicative structure on the lattice, wherein adjacent vertices are related strictly by group-valued fields. The result can be viewed as a discrete realization of the integral formalism \cite{yang_integral_1974} of early gauge theory.

By virtue of its manifest gauge invariance, Poincar\'e $\mOne$-loop theory improves upon Regge calculus as a classical, discrete theory of gravity. Its continuum limit, as derived in Eqs.~(\ref{continuum1Loop5VT})-(\ref{schematicFieldEq}), recovers the field equations of the YM-track of Poincar\'e gauge theory \cite{popov_theory_1975,popov_einstein_1976,tseytlin_poincare_1982,aldrovandi_complete_1984,aldrovandi_natural_1986,aldrovandi_quantization_1988}. Despite the YM-track exhibiting many promising features of a gravitational theory \cite{aldrovandi_complete_1984}, the incompatibility of its field equations with Lagrangian mechanics has led some of its investigators to view the YM-track with disfavor \cite{tseytlin_poincare_1982,aldrovandi_natural_1986,aldrovandi_quantization_1988}. Poincar\'e $\mOne$-loop theory addresses some of the concerns raised in this prior work, as we now discuss.

First, the determination of matter couplings in the YM-track has not been well understood; for example, the interpretation of $T$ in Eq.~(\ref{schematicFieldEq}) as an energy-momentum has been in doubt \cite{tseytlin_poincare_1982}. $\mOne$-loop theory addresses this issue by defining a new formalism that explicitly defines the properties of a matter current and its coupling to a gauge field. In Eq.~(\ref{defineTheCurrent}), this formalism was shown to uniquely determine $\mJ_a$, the $\mmp$-valued current of the 5-vector field. Second, from a more philosophical point of view, authors of the YM-track caution generally against its failure to derive from a variational principle \cite{tseytlin_poincare_1982,aldrovandi_natural_1986}. However, the crucial use of Cartan geodesics in the $\mOne$-loop formalism lends it a variational character, even absent a Lagrangian.

Lastly, some authors of the YM-track note that, although it has not made unphysical predictions of classical gravitational dynamics, its lack of a Lagrangian complicates its quantization via a path integral approach. This difficulty is understood to render the YM-track unsuitable as a quantum theory of gravity \cite{aldrovandi_quantization_1988}.

In the present work, we have aspired to a more modest goal: a classical, computable, and physically sensible algorithm that preserves Poincar\'e symmetry in a discrete universe. $\mOne$-loop theory is essentially a computable physical theory stripped of all considerations except symmetry principles. Its constituents---holonomies and Noether currents---arise directly as representations of a symmetry group, with as little additional structure as possible. With this spare framework, $\mOne$-loop theory demonstrates that there is no essential conflict between discrete space-time and Poincar\'e symmetry. An algorithm for gravitational simulations has thus been developed that, in principle, covariantly conserves energy and momentum to machine precision.

In future work, we shall consider the physical effects of discreteness on the gravitational dynamics of Poincar\'e $\mOne$-loop theory. We shall also explore $\mOne$-loop theories for more inclusive gauge groups, such as ${G=\mP\times U(1)}$, and for fermionic $\mP$-representations, whose spin and orbital angular momenta can both be expected to appear in the matter current.

\section{Acknowledgments}

Thank you to Eugene Kur for illuminating discussions, and to Professor Nathaniel Fisch for his encouragement of this effort. A.S.G. acknowledges the generous support of the Princeton University Charlotte Elizabeth Procter Fellowship. This research was further supported by the U.S. Department of Energy (DE-AC02-09CH11466).

\bibliography{allrefs.bib}

\end{document}